# Methodology for Business Intelligence Solutions in Internet Banking Companies

Alex Escalante Viteri [a,*], Javier Gamboa Cruzado [a], Leonidas Asto Huaman [b]

[a] *Faculty of Systems Engineering, Postgraduate Unit, Universidad Nacional Mayor de San Marcos, Cercado, Lima, 15081, Perú*
[b] *Professional School of Industrial Engineering, Universidad Nacional Autónoma de Tayacaja, José Olaya, Pampas, 0956, Perú*
*Corresponding author: \*alex.escalante@unmsm.edu.pe*

*Abstract*— Business intelligence in the banking industry has been studied extensively in the last decade; however, business executives still do not perceive efficiency in the decision-making process since the management and treatment of information are very time-consuming for the deliverer, generating costs in the process. On the other hand, there is no formal methodology for developing business intelligence solutions in this sector. This work aims to optimize decision-making in a business unit that works with internet banking companies, reducing the time, the number of people, and the costs involved in decision-making. To meet the objective, basic and applied research was conducted. The basic research allowed the construction of a new methodology from a study of critical success factors and approaches from the business intelligence literature. The applied research involved the implementation of a business intelligence solution applying the new methodology in a pre-experimental study. Thirty decision-making processes were analyzed using pre-test and post-test data. Tools such as a stopwatch and observation were used to collect and record data on time spent, the number of people, and the decision-making costs. This information was processed in the specialized Minitab18 statistical software, which allowed the observation and confirmation of relevant results regarding time reduction, the number of people, and the costs generated. Therefore, it was concluded that the business intelligence solution, applying the new methodology, optimized decision making in the business unit that works with internet banking for companies.

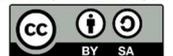

## I. INTRODUCTION

Business intelligence solutions are part of the global financial system, which is in a new stage of development, characterized by the introduction of information and communication technologies in all financial areas [1]. In this sense, decision-making in the internet banking business is especially significant for banks because legal clients or companies that use internet banking have high performance and liquidity in banking management [2]. They can also adjust their business strategies to improve the disposition of their consumers [3]. According to Yiu et al. [4], the implementation of business intelligence systems leads to greater operational capacity, particularly for large high-tech companies with high technological intensity and considering the strategic impact on business sustainability [5]. Therefore, it is important to apply appropriate methodologies to develop business intelligence solutions that help optimize decision-making in Peru because the use of virtual channels in the Peruvian financial sector has been increasing significantly. According to Morisaki [6], during the first months of 2019, 430.5 million transactions were non-cash payments, and 19.79% (85.2 million) of transactions were made through virtual channels (internet banking, mobile banking, and commercial internet). Due to the COVID-19 pandemic, in May 2020, virtual channels surpassed face-to-face channels for the first time [7].

To develop business intelligence solutions, conceptual and theoretical frameworks of critical success factors have been proposed. These need to be evaluated in organizations [8], [9], [10] as well as concerning the factors that influence the selection of software tools [11]. For example, Ranjbarfard and Hatami [12] proposed a relationship between critical success factors and business intelligence methodologies that are more representative. The results revealed eleven critical success factors to consider in a business intelligence project. Adeyelure et al. [13] studied a proposed framework for deploying mobile business intelligence in small and medium-sized enterprises in developing countries. The factors of mobile business intelligence in another study were investigated through textual analysis [14]. Rezaie et al. [15] measured the key factors affecting the business intelligence implementation process and important effectiveness criteria for business intelligence in the Iranian banking industry.

Another factor that is of interest in defining a new methodology is the adoption of business intelligence systems. Ahmad et al. [16] and Ain et al. [17] attempted to reduce this gap through a systematic review of the literature. The adoption of business intelligence at the internet service level is another important indicator to take into account [18]. Rouhani et al. [19] conducted an in-depth analysis to understand the critical factors that affect the decision to adopt business intelligence in the context of the banking and financial industry.

*A. Business Intelligence Solutions in Banking*

Various business intelligence solutions could apply a new methodology. This would entail considering the key factors that influence developments [20] and agile values that contribute to the success of business intelligence solutions [21]. Gonzáles-Carrasco et al. [22] found that increasing the use of big data and artificial intelligence techniques improved the customer experience.

Predictive models have been built to study, for example, the client's retirement journey and to create a model for predicting churn [23]. On the other hand, financial and insurance services based on the use of IoT applications have also been examined [24].

A business intelligence solution must conform to a methodology to meet the minimum requirements of the business user. For example, Massardi et al. [25] designed a business intelligence application that adhered to the financial ratios required by the user to analyze the financial condition of rural banks. Another example is a business intelligence solution that described the importance of data selection over storage and OLAP development. This can help managers to make better decisions [26]. Therefore, user requirements are important for the organizational performance of banks since the effective adoption of business intelligence systems depends on them. This was evident in the case of a study carried out in the universal banks of Ghana [27].

Business intelligence solutions can be especially beneficial in decision-making if developed properly to gain competitive advantages and productivity. Al-Maaitah [28], in his study, identified the role of business intelligence skills (managerial skills, technical skills, and cultural skills) in the organizational capabilities of Jordanian banks (process improvement, innovations, flexibility, and agility). According to the author, business intelligence competencies have a significant impact on the organizational capabilities of Jordanian banks. In Poland, business intelligence solutions were used in a review of productivity, quality, profitability, and debt [29].

Mortezaei et al. [30], however, investigated the role of business intelligence competence in improving the customer relationship management process. They developed a conceptual model that encompassed different dimensions of business intelligence competency and customer relationship management processes.

It is important to mention that data mining and neural networks are recommended in banking and business intelligence. For example, studies have applied algorithms based on artificial colonies of ants [31]. Alzeaideen [32], however, developed an artificial neural network model as a decision support system for evaluating credit approval in Jordanian commercial banks. Finally, it is important to mention blockchain technology, and thas a prospect for application in the future financial industry [33], [34].

*B. Approaches and Critical Success Factors*

According to Trieu [35], much of the research on business intelligence has examined the ability of business intelligence solutions to help organizations address challenges and opportunities. Nevertheless, the literature is fragmented and lacks an overarching framework to systematically integrate findings and guide research.

Critical success factors and approaches were very important in the present study. According to the research problem, these are starting points to ensure the greatest number of approaches. We can identify the critical factors to ensure a successful project based on this. Table I shows the identified approaches with relevance to the research problem.

TABLE I
BUSINESS INTELLIGENCE APPROACHES

| Author | Description of approaches |
|---|---|
| [36], [37] | *Plan-oriented approach or requirement-oriented approach*: This is considered to be a traditional approach; however, it is hard for users to define and explain how they make their decisions. *Focus on data management*: This approach focuses on data: how they are structured, who uses them, and how they use them. The data drive the process. *Focus on value chain data*: This approach is an evolution of the "data management" approach concentrating on the data that will generate the greatest value for the business. *Process-based approach*: This approach is based on the analysis of business processes, the information they generate, and the information they consume. *Event-driven approach*: This approach proposes to divide the business processes into three points of view, data, function, and organization, each of which is connected through events. *Object-oriented approach*: In this approach, both objects and processes have the same importance from a decisional point of view and should be treated in the same way. *Joint approach*: The main idea of this approach is that the organization is a matrix of processes with different information needs, but where they come together is where the greatest effort is required. *Goal-oriented approach*: This approach focuses on the objective of the strategic processes of the organization and is based on the analysis of the interaction of customers and users to achieve this objective. *Model-based approach*: This approach intends to build a bridge between the business and the IT department to provide the basis for developing quick solutions that evolve easily and flexibly. *Adaptive business approach*: This focuses on the problems that a business has to solve to adapt to market changes and on the data that we have for this. |
| [38] | *Demand-driven or prototype-driven user-based approach*: This approach is based on methodologies oriented towards the making of prototypes to obtain sufficiently precise results. |
| [39] | *Three-pronged management approach*: It is necessary to commit to a combination of the best ideas of each of the objective, data, and user management methodologies, creating triple management since these three approaches are considered to be perfectly compatible. |
| [40] | *Agile approach*: This approach can be considered novel in the area of software engineering and even more so in the area of business intelligence. |

Studies on techniques to choose the best strategic alternative are compelling research [41]. These techniques allow us to consider different alternatives to ensure a successful project. The critical success factors of business intelligence were analyzed to select the best options for a successful project. Critical success factors played an important role and allowed a previous analysis of methodological approaches to mature and offer more efficient business intelligence solutions. Table II shows the critical success factors selected from the literature relating to our research problem.

TABLE II
CRITICAL SUCCESS FACTORS OF BUSINESS INTELLIGENCE

| Author | Description of critical success factors |
|---|---|
| [42], [43] | Include user participation in the definition of the level of service and the requirements |
| | Define the quality plan |
| | Choose the ETL (extract, transform, and load) tool to use |
| | Preferably carry out incremental data loads |
| | Carefully choose the development platform and the appropriate database system |
| | Carry out data reconciliation processes |
| | Periodically review and modify the planning |
| | Provide user support |
| [44], [45], [46] | Sponsorship of the project |
| | Management of user expectations |
| | Use of prototypes |
| | Quick result search (quick win) |
| | Choose a measurable organizational problem |
| | Modeling and design of the "data warehouse." |
| | Selection of the appropriate business case |
| | Alignment with the organizational strategy |
| | Careful selection of tools |
| | End users' involvement |
| [47], [48], [49] | Support for the management of the organization |
| | Existence of a project leader |
| | Adequate use of resources |
| | Participation of the end-user |
| | Team with adequate skills |
| | Have adequate data sources |
| | Consider the information and its analysis as part of the organization's culture |
| | Alignment with the organization's strategy |
| | Effective BI management and control |
| | Management of organizational change |
| [50] | Initiative linked to business needs |
| | Existence of management sponsorship |
| | Cross-organizational project |
| | QA control |
| | The flexibility of the data model |
| | Data-oriented management |
| | Automatic data extraction process |
| | Knowledge |
| | Experience |
| [51] | Make incremental changes |
| | Adaptive system construction |
| | Manage user expectations |
| | Mixed team involving technicians and end-users |
| | Direct contact with the organization and the business |
| | Avoid chasing perfection |
| | Transmit knowledge in subcontracted projects |
| | Use of standards |
| | Take advantage of the experience of the team members |
| | End-user support |
| [52] | The centralization of data in a "data warehouse" and its division into several "data marts" allow fast and reliable access to the requested information |
| | The definition of standard lists for all users favors the exchange of information between departments in a clearer and more consistent way |
| | Some predefined report templates have to be implemented to provide decision makers with the functionality to add or remove particular items and create specific reports |
| | A team responsible for aligning standard reporting specifications with local needs and facilitating the execution of the BI project is necessary |
| | There must be strong commitment from the management to resolve any conflict and manage changes that occur during the development of the project |
| | Integration of "Six Sigma" techniques into the organization's IT infrastructure contributes to a robust BI system |
| | IT infrastructure has to focus on a single platform provided by well-known vendors |
| | Consideration of the culture of the organization |
| | Focus on data management |
| | Level of scalability and flexibility of the project and the solution |
| [53] | Senior management's support for the project |
| | Adequate resources |
| | Committed support from the organization |
| | Formal user participation throughout the entire project |
| | Support, education, and training |
| | Established and agreed business case |
| | Strategic BI vision integrated with the company initiatives |
| | Clearly defined scope of the project |
| | Adoption of an incremental results approach |
| | Project oriented to achieve quick results (quick wins) |
| | Team with the perfect combination of capabilities |
| | Participation of external consultancy in the initial phases of the project |
| | Experience in the business domain |
| | Multifunctional team |
| | Stable data provider systems |
| | Strategic, scalable, and extensible technical environment |
| | Use a prototype as proof of concept |
| | Quality data sources |
| | Common metrics and classifications established by the organization |
| | Scalable metadata model |

This research aims to optimize decision making in a business unit that works with internet banking companies through basic and applied research. Section II shows how the basic and applied research was conducted. Regarding the basic research, it describes how the new proposed methodology was built, and, in relation to the applied research, it shows how the tools and techniques were used during the pre-experimental research. Section III contains the results of the data analysis using Minitab18 statistical software and the hypotheses presented in section II. Finally, section IV reveals the conclusion based on the objective of the investigation.

## II. Materials and Method

This study was conducted in a bank business unit that works with internet banking companies. It identified time problems, the number of people, and the costs generated in decision making. Accordingly, basic research was conducted to propose a new business intelligence methodology. Applied research was then conducted that used the new methodology to implement a business intelligence solution to solve the problem. Figure 1 shows the method used in the research.

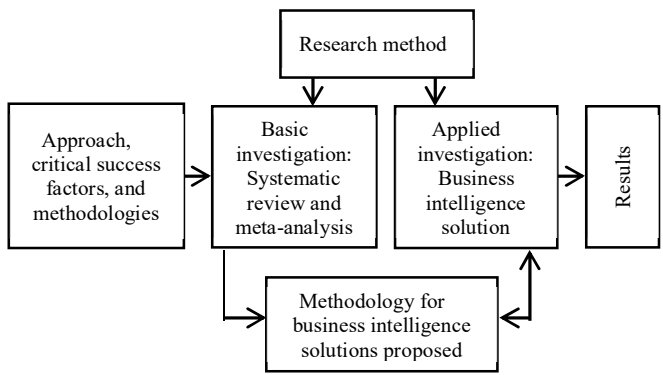

Fig. 1 Research method

### A. Basic Investigation Method

First, a systematic review of the literature was conducted, filtering four business intelligence methodologies with thirteen representative approaches. The analysis technique was used to determine which methodology best suits a certain business intelligence approach. Table III shows the methodologies with the highest relationship score: Ralph Kimball (RK), DWEP (DW), and SAS Rapid (SR). Bill Immon (BI) and Hephaestus (HF) were eliminated.

TABLE III
BUSINESS INTELLIGENCE APPROACHES AND METHODOLOGIES

| Author | BI approaches | RK | BI | HF | DW | SR |
|---|---|---|---|---|---|---|
| [36], [37] | The plan-oriented approach or requirement-oriented approach | X | X | | | |
| [36], [37] | Focus on data management | | X | | | X |
| [36], [37] | Focus on value chain data | | X | | | X |
| [36], [37] | Process-based approach | | X | | X | |
| [36], [37] | Event-driven approach | | X | | | X |
| [36], [37] | Object-oriented approach | X | | | | X |
| [36], [37] | Joint approach | | X | | | |
| [36], [37] | Goal-oriented approach | X | | X | X | X |
| [36], [37] | Model-based approach | X | | | X | |
| [36], [37] | Adaptive business approach | X | | X | X | X |
| [38] | Demand-driven or prototype-driven user-based | | | | | |
| [39] | Three-pronged management approach (objective, data, and user management) | X | | X | X | X |
| [40] | Agile approach | X | | | X | X |

Second, a meta-analysis of more representative critical success factors was conducted. Critical success factors were weighted, ranked, and related to the methodologies that obtained the most approaches in the first analysis.

The QSPM technique was used. The Quantitative Strategic Planning Matrix (QSPM) is a high-level strategic management approach for evaluating possible strategies, and it provides an analytical method for comparing feasible alternative actions. This technique assigns weights, classifications, and scores. A weight (b) and a classification (c) are given to each critical success factor, whereby multiplying b □ c will obtain a score for each methodology, which will be considered to be more attractive for a successful project. In this case, it will be in accordance with the critical success factor proposed by each author in the literature. David et al. [41] stated, "The criterion for the quantitative matrix is to determine the relative attractiveness of viable relative actions".

Each critical success factor was assigned a weight: weight = 0.0 = unimportant, weight = 0.1 = very important. The sum must always be equal to 1. A classification is then assigned to each strategic element as a degree of attraction (in this case, the strategically selected methodologies): 1 = not attractive, 2 = somewhat attractive, 3 = quite attractive, and 4 = very attractive.

The total score of the degree of attraction was obtained by multiplying the values of the ranking by weight; the total scores indicated the degree of attraction of each strategy. In Table IV, we can see the values found.

TABLE IV
SUMMARY OF THE META-ANALYSIS OF CRITICAL SUCCESS FACTORS

| Methodologies selected | | Ralph Kimball | | DWEP | | SAS RAPID | |
|---|---|---|---|---|---|---|---|
| FCE according to author (a) | Average weight (b) | Average rating (c) | Total score (d) = (b * c) | Average rating (e) | Total score V(f) = (b * e) | Average rating (g) | Total score (h) = (b * g) |
| [42], [43] | 1.0 | 2.8 | 2.86 | 3.1 | 3.12 | 2.5 | 2.44 |
| [44], [45], [46] | 1.0 | 3.0 | 3.13 | 2.7 | 2.62 | 2.2 | 2.40 |
| [47], [48], [49] | 1.0 | 2.5 | 2.80 | 2.3 | 2.36 | 2.2 | 2.10 |
| [50] | 1.0 | 2.8 | 3.08 | 2.9 | 2.88 | 2.4 | 2.24 |
| [51] | 1.0 | 3.0 | 3.36 | 2.3 | 2.21 | 2.5 | 2.75 |
| [52] | 1.0 | 2.6 | 2.67 | 2.6 | 2.68 | 2.1 | 2.25 |
| [43], [53] | 1.0 | 2.9 | 3.0 | 2.3 | 2.06 | 2.7 | 2.74 |
| [43], [53] | 1.0 | 2.4 | 2.52 | 2.4 | 2.6 | 2.5 | 2.5 |
| [43], [53] | 1.0 | 2.8 | 2.94 | 2.9 | 3.22 | 1.9 | 1.8 |
| [43], [53] | 1.0 | 2.8 | 3.04 | 3.1 | 3.4 | 2.5 | 2.62 |
| Total | | | 29.4 | | 27.15 | | 23.84 |

Taking into account the highest scores of the meta-analysis, a new methodology was proposed. The most relevant concepts of the Ralph Kimball and DWEP methodologies were considered. Figure 2 shows the phases of the new proposed methodology used in applied research.

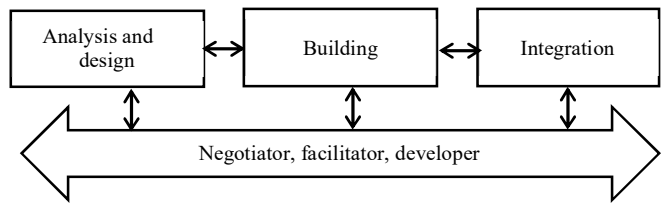

Fig. 2 Methodology proposed

This new methodology allows the efficient implementation of a business intelligence solution since it maintains its three development phases through communication between three actors who manage the information transversally in all the phases of the project.

*1) The Analysis and Design Phase*: This phase was used to obtain business value requirements through the method of interviews and observation. It allowed the design of the use cases, the architecture of the solution, and the entire flow of data input and output.

*2) The Construction Phase*: This phase facilitated the development of databases and the execution of ETL processes and automated dimensional cubes. It avoided impediments due to new changes. These changes were facilitated and accepted as a principle of agility.

*3) The Integration Phase*: This phase allowed for incremental deliverables. The methodology required the dashboards to be the business acceptance criteria, and it allowed for comprehensive tests that were verified by the same business. Finally, the solution was implemented and investigated within a culture of development and operations.

*B. Applied Research Method*

The new methodology was applied by developing a business intelligence solution to optimize decision-making in internet banking. The following variables and indicators were considered:

*1) Independent Variable*: A business intelligence solution applying a new methodology.

TABLE V
OPERATIONALIZATION OF THE INDEPENDENT VARIABLE

| Indicator | Index |
|---|---|
| Presence, absence | No, yes |

When the answer is NO, it is because the new methodology was not applied during the development of a business intelligence solution. The problem is still in its current situation. When it is YES, it refers to the new methodology being applied to the development of a business intelligence solution, which is expected to obtain better results.

*2) Dependent Variable*: Decision making in internet banking companies. Table VI shows the operationalization of the indicators of the dependent variable.

TABLE VI
OPERATIONALIZATION OF THE DEPENDENT VARIABLE

| Dimension | Indicator | Index | Unit of measurement | Formula | Method |
|---|---|---|---|---|---|
| Time | Time spent in each decision making process | [20–120] | Minutes | T = (TI1 + TIn)/NS | Direct observation |
| People | Number of people in each decision making process | [2–5] | # of people used | P = (PN + PS) | Direct observation |
| Cost | Cost per hour generated in each decision-making process | [31,25] | Peruvian soles | C = (CH ☐TTD/60) ☐ NP | Manual review |

Formula legend:

*Time*:
T = time invested in the process
TI1 + TIn = time spent according to 1 + n elements of the process
NS = number of outputs of a process

*Persons*:
PN = people involved in business
PS = people involved in systems

*Cost*:
CH = cost per hour
TTD = time in minutes in the decision-making process
NP = number of people involved

*3) Pre-experimental Research*: The indicated formulas were applied in a pre-experimental research design because they worked with only one research group.

TABLE VII
EXPERIMENTAL DESIGN NOTATION

| RG | 01 | X | 02 |
|---|---|---|---|
| Experimental group | Pre-test measurement | Experimental treatment | **Post-test measurement** |

where:

R = random group formation (random choice of decision-making processes)
G = experimental group (decision-making processes)
01 = pre-test measurement (values found before the experimental stimulus)
X = experimental treatment (business intelligence solution applying the new methodology)
02 = post-test measurement (values found after applying the experimental stimulus)

The decision-making processes of internet banking companies worldwide were considered as a population; however, as it was not possible to quantify all the processes:

*N = indeterminate*

The decision-making processes of internet banking companies were taken as a sample.

*n = 30 decision-making processes carried out*

Direct observation was used as a technique and the stopwatch as a research tool to measure the indicators – time, number of people, and cost of the decision-making processes – in the pre-test and post-test of the data. Each process was analyzed in detail according to the indicators, and notes were taken to perform the calculations and obtain the results.

*4) Statement of the Hypotheses*: The parameter studied was the average (µ) of the indicators: time, the number of people, and the cost of the decision making of the internet banking companies. Therefore, the following hypotheses were stated:

- $H_1$: If a business intelligence solution is implemented with a new methodology, the time spent on decision making for internet banking companies is reduced.
- $H_2$: If a business intelligence solution is implemented with a new methodology, the number of people participating in internet banking companies' decision-making is reduced.
- $H_3$: If a business intelligence solution is implemented with a new methodology, the cost of decision making for internet banking companies is reduced.

To contrast the hypotheses, the following solution was proposed for each of the indicators:

$\mu_1$ = mean ($H_1$, $H_2$, $H_3$) of decision making in the pre-test
$\mu_2$ = mean ($H_1$, $H_2$, $H_3$) of decision making in the post-test
where:

$H_0$: $\mu_1 \leq \mu_2$
$H_a$: $\mu_1 > \mu_2$

Finally, the hypotheses were confirmed using the specialized software Minitab18. Data normality analysis, descriptive statistics analysis, and hypothesis contrast analysis were performed for statistical decisions.

### III. RESULTS AND DISCUSSION

*A. Reduction of Time, Number of People, and Cost*

The effect of applying the business intelligence solution using the new proposed methodology had significant results. It reduced the time and the number of people involved and the costs generated in the decision-making process. Thirty decision-making processes were observed. The pre-test results determined the time that each process takes according to the tasks performed by the number of people involved and the cost generated by each person's work. In the subsequent test, a significant reduction of the indicators was observed. The business executive used the business intelligence solution to make decisions in a shorter time, using fewer people and generating lower costs. This new result was important to determine the time, number of people involved, and the cost generated in a new decision-making process.

Table VIII shows that 100% of the data on the decision making in the post-test are lower than the average of the data in the pre-test. It was observed that 67% of the decision-making time in the post-test is less than the average time. For the number of people, it is 63% less, and, for the cost, it is 70% less.

TABLE VIII
DIRECT OBSERVATION RESULTS

| No. | $I_1$: Time | | $I_2$: Number of people | | $I_3$: Cost | |
|---|---|---|---|---|---|---|
| | Pre-test | Post-test | Pre-test | Post-test | Pre-test | Post-test |
| 1 | 63.33 | 1.77 | 3 | 0 | 84.87 | 0.42 |
| 2 | 35.00 | 3.73 | 3 | 1 | 58.34 | 1.62 |
| 3 | 21.67 | 1.41 | 4 | 1 | 47.00 | 0.44 |
| 4 | 20.00 | 1.69 | 3 | 1 | 29.69 | 0.47 |
| 5 | 71.67 | 18.33 | 3 | 1 | 100.06 | 9.17 |
| 6 | 90.00 | 26.67 | 4 | 0 | 173.87 | -0.73 |
| 7 | 76.67 | 15.00 | 3 | 1 | 104.90 | 6.04 |
| 8 | 106.67 | 28.33 | 2 | 1 | 120.30 | 17.72 |
| 9 | 101.67 | 25.00 | 3 | 1 | 183.49 | 10.68 |
| 10 | 76.67 | 10.00 | 2 | 1 | 97.52 | 6.73 |
| 11 | 65.00 | 7.67 | 3 | 2 | 86.01 | 7.06 |
| 12 | 68.33 | 20.00 | 3 | 1 | 96.62 | 9.54 |
| 13 | 65.67 | 25.00 | 4 | 1 | 145.99 | 18.21 |
| 14 | 38.33 | 7.67 | 2 | 0 | 37.25 | 1.50 |
| 15 | 38.33 | 3.33 | 2 | 2 | 42.60 | 3.00 |
| 16 | 18.33 | 0.60 | 3 | 1 | 30.20 | 0.43 |
| 17 | 25.00 | 0.43 | 3 | 1 | 38.08 | 0.17 |
| 18 | 43.33 | 1.50 | 3 | 1 | 57.37 | 0.50 |
| 19 | 61.67 | 0.12 | 5 | 2 | 168.20 | 0.11 |
| 20 | 52.33 | 0.08 | 4 | 1 | 97.67 | 0.03 |
| 21 | 25.00 | 1.67 | 4 | 0 | 48.42 | 0.21 |
| 22 | 46.67 | 0.32 | 3 | 1 | 65.08 | 0.24 |
| 23 | 81.67 | 7.67 | 4 | 2 | 188.21 | 9.64 |
| 24 | 78.33 | 11.67 | 4 | 0 | 161.93 | 2.00 |
| 25 | 80.00 | 15.00 | 2 | 0 | 68.10 | -1.03 |
| 26 | 35.00 | 0.10 | 5 | 2 | 87.81 | 0.09 |
| 27 | 76.67 | 7.67 | 2 | 1 | 92.98 | 2.43 |
| 28 | 55.00 | 0.04 | 4 | 1 | 102.29 | 0.03 |
| 29 | 76.67 | 0.08 | 4 | 0 | 154.56 | 0.01 |
| 30 | 90.00 | 0.38 | 2 | 1 | 81.60 | 0.11 |
| µ | **59.49** | **8.10** | **3** | **1** | **95.03** | **3.56** |
| Nro. <= µ | 30 | 20 | 30 | 19 | 30 | 21 |
| % | **100** | **67** | **100** | **63** | **100** | **70** |

The direct observation method using the stopwatch as a recording tool is useful and allows empirical knowledge to be obtained to understand reality. Nevertheless, a data bias could have arisen when recording the information, so it can be understood as subjective. Therefore, the Minitab18 software was applied as a standardized method to supply the results with statistical evidence.

*1) Descriptive Statistics Results*: *According to the results of the "Anderson Darling" normality test, the AD and p-value are > α (0.05); therefore, the data normality was confirmed for analysis.* It was observed that, with a confidence level of 95%, the mean and the standard deviation revealed normal results concerning the data of the research indicators.

Table IX shows the results of the descriptive statistics according to the Minitab18 software.

TABLE IX
DESCRIPTIVE STATISTICS RESULTS

| Sample | N | Mean | Stand. dev. | AD | p-value |
|---|---|---|---|---|---|
| $I_1$: Pre-test – time | 30 | 59.49 | 25.11 | 0.460 | 0.243 |
| $I_1$: Post-test – time | | 1.350 | 0.5568 | 0.637 | 0.088 |
| $I_2$: Pre-test – number of people | 30 | 3.120 | 0.9199 | 0.358 | 0.431 |
| $I_2$: Post-test – number of people | | 0.9368 | 0.6028 | 0.431 | 0.287 |
| $I_3$: Pre-test – cost | 30 | 95.03 | 47.93 | 0.721 | 0.054 |
| $I_3$: Post-test – cost | | -0.1066 | 2.051 | 0.378 | 0.386 |

It was observed that the pre-test values are higher than the post-test values. This provides evidence that the time was reduced from 59.49 to 1.35 minutes. The number of people was reduced from 3 to 1, and the average cost generated per hour was reduced from 95.03 to -0.1066. Therefore, the process of decision making was significantly reduced.

*2) Hypothesis Testing Results*: The results of the statistical values of the differences in the samples in the pre-test and the post-test are shown in Table X. The t- and p-values are verified in Figure 3 to decide the regions of acceptance and rejection.

TABLE X
STATISTICAL VALUES OF ACCEPTANCE AND REJECTION

| Sample | N | t-value | p-value |
|---|---|---|---|
| $I_1$: Pre-test – time / $I_1$: Post-test – time | 30 | 12.84 | 0.000 |
| $I_2$: Pre-test – number of people / $I_2$: Post-test – number of people | 30 | 12.92 | 0.000 |
| $I_3$: Pre-test – cost / $I_3$: Post-test – cost | 30 | 10.95 | 0.000 |

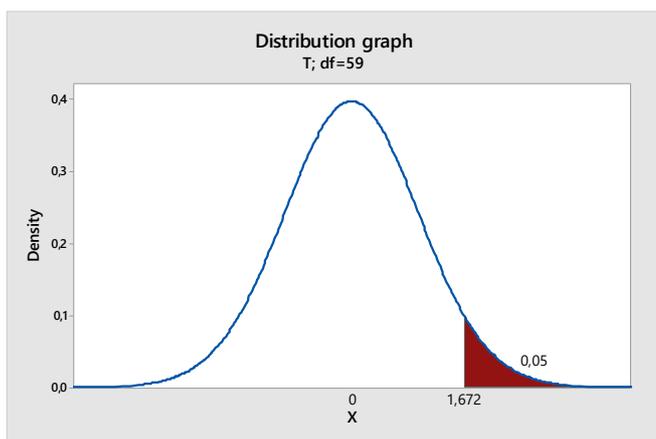

Fig. 3 Statistical acceptance limit

With a degree of freedom (n-1) = 59 for the two samples, a critical value of 1.672 was obtained, limiting the acceptance zone to 95% and producing a right-tailed rejection of 0.05. According to the result of the t-value calculated from the samples in Table X, it is within the rejection zone, and the result of the value p = 0.000 < α = 0.05 shows sufficient evidence to consider the test of the hypothesis to be significant.

*3) Statistical Decision Result*: $H_0$ is rejected: $\mu_1 \leq \mu_2$ and $H_a$ is accepted: $\mu_1 > \mu_2$. It is concluded that, at the significance level of 5%, the null hypothesis $H_0$ is rejected. The indicators "time", "number of people", and "cost" in the pre-test are less than or equal to those in the post-test. It is considered that there is sufficient statistical evidence to accept the alternative hypothesis. The decision-making indicators are higher in the pre-test than in the post-test. Therefore, it is resolved that the hypothesis indicators $H_1$, $H_2$, and $H_3$ are true with significant results.

*B. The Effect on Internet Banking Companies*

If business executives make decisions in the shortest possible time, using fewer people, and involving lower costs, they will be more productive. Performing data queries using a business intelligence solution that was developed with an effective methodology will allow greater agility in decision making. Business units will be able to improve their strategies and offer high-quality services with greater productivity on their internet banking platforms [3]. Customers will then have a high operational capacity [4] due to the offer of products that they can receive.

The new methodology for business intelligence solutions in internet banking returned significant results. These indicate greater business productivity and continued participation in the virtual channel market [7].

It is important to highlight the adoption of business intelligence methodologies for any type of information technology to make further improvements [17]. On the other hand, artificial intelligence and block technology are considered to be prospects for the future of banking [31], [33]. In this study, the results that were presented, with a reduction from 59.49 to 1.350 minutes in the work per process, are extremely significant for decision-making in the internet banking business unit. This will result in the project's development requests and product offerings being made more frequently. Business executives will be able to avoid opportunity costs in the business. On the other hand, the reduction from three to one person to generate decision-making reports can eliminate dependencies and gaps, enabling information to be obtained more efficiently. Finally, reducing people will reduce costs, and this reduction can be considered savings for investment in other opportunities in the business of internet banking companies.

IV. CONCLUSION

Decision-making was optimized in a financial sector business unit that offers internet banking to business clients. This optimization was achieved by implementing a business intelligence solution applying a new business intelligence solution methodology. The business unit will be able to make better decisions in less time, with fewer people, and with lower costs.

ACKNOWLEDGMENT

We thank the Universidad Nacional Mayor de San Marcos for accepting the research for a postgraduate thesis as well as the financial entity in Peru for allowing us to collect the necessary information.